\documentstyle[aas2pp4]{article}
\begin{document}
 \title{SUPERMASSIVE BLACK HOLE BINARIES AS GALACTIC BLENDERS}
\author{HENRY E. KANDRUP}
\affil{Department of Astronomy, Department of Physics, and Institute for
Fundamental Theory, University of Florida, Gainesville, FL 32611}

\author{IOANNIS V. SIDERIS} 
\affil{Department of Physics, Northern Illinois University, DeKalb, IL
60115}
 
\author{BAL{\v S}A TERZI{\'C}}
\affil{Department of Astronomy, University of Florida, Gainesville, FL 32611}

\and

\author{COURTLANDT L. BOHN} 
\affil{Department of Physics, Northern Illinois University, DeKalb, IL
60115 and Fermilab, Batavia, IL 60510}


\begin{abstract}
This paper focuses on the dynamical implications of close supermassive black
hole binaries both as an example of resonant phase mixing and as a potential
explanation of inversions and other anomalous features observed in the
luminosity profiles of some
elliptical galaxies. The presence of a binary comprised of black holes
executing nearly periodic orbits leads to the possibility of a broad resonant
coupling between the black holes and various stars in the galaxy. This can
result in efficient chaotic phase mixing and, in many cases, systematic
increases in the energies of stars and their consequent transport towards
larger radii. Allowing for a supermassive black hole binary
with plausible parameter values near the center of a spherical, or nearly
spherical, galaxy characterised initially by a Nuker density profile enables
one to reproduce in considerable detail the central surface brightness
distributions of such galaxies as {\em NGC} 3706.
\end{abstract}


\section{Introduction and Motivation}

Understanding the dynamical implications of a supermassive black hole binary
near the center of a galaxy is important {both} because of the insights
the problem can shed on physical processes associated with a time-dependent
potential {and} because, even if it itself is not resolvable
observationally, the binary can have directly observable effects.

As is well known to nonlinear dynamicists, a time-dependent potential can
induce significant amounts of time-dependent {\em transient chaos}, an interval
during which orbits exhibit an exponentially sensitive dependence on initial
conditions, and resonant couplings
between the natural frequencies of the time-dependent potential and the
frequencies of the chaotic orbits can trigger efficient {\em resonant phase
mixing} (Kandrup, Vass \& Sideris 2003). Like `ordinary' chaotic phase mixing
({\em e.g.,} Kandrup \& Mahon 1994, Merritt \& Valluri 1996), this resonant
mixing can facilitate a rapid shuffling of orbits on different constant
energy hypersurfaces. Even more importantly, however, because
the potential is time-dependent the energies of individual orbits are not
conserved, so that resonant mixing can also facilitate a shuffling of energies
between different constant energy hypersurfaces.

For this reason, resonant phase mixing has important implications for
collective relaxation in nearly collisionless systems (Kandrup 2003),
{\em e.g.,} holding forth the prospect of explaining from first principles
the striking efficacy of violent relaxation (Lynden-Bell 1967) 
found in simulations and inferred from 
observations
(see, {\em e.g.,} Bertin 2000).
That large scale collective oscillations could trigger very efficient
violent relaxation has been shown in the context of one simple
model, namely orbits of stars in a Plummer sphere subjected to a
systematic time-dependence which eventually damps (Kandrup, Vass \&
Sideris 2003). The binary black hole problem provides a complementary
example of how 
smaller scale time-dependences can also have a surprisingly large effect.

The binary black hole problem is also interesting because the 
binary can have directly observable consequences. The fact that energy is
not conserved implies the possibility of readjustments in the density profile
of stars near the center of a galaxy. In many cases this energy
nonconservation means that, on the average, stars near the center gain energy,
which implies a systematic transport of luminous matter near the black
holes out to larger radii. To the extent, however, that mass traces light, such
changes in the density distribution translate into predicted
changes in the observed surface brightness distribution because
of the presence of such a binary.

In particular, for reasonable
choices of black hole masses and orbital parameters, the binary
can actually cause an `inversion' in the surface brightness profile,
so that surface brightness is no longer a monotonically decreasing function of
distance from the center. Indeed, the simplest models which one might envision
are adequate to reproduce distinctive features observed in  the 
brightness distributions of such galaxies as NGC 3706, as
reported in Lauer {\em et al} (2002).

The first half of this paper, Section 2, considers the binary black hole 
problem as an example of how a time-dependent potential can facilitate 
efficient phase mixing in a galaxy. Attention focuses on two sets of 
models, namely the pedagogical example of a constant density ellipsoid,
corresponding to an anisotropic oscillator potential, 
and more realistic cuspy density profiles 
consistent with what have been inferred from high resolution photometry 
({\em e.g.,} Lauer {\em et al} 1995).

One important issue here involves determining as a function
of amplitude ({\em i.e.,} black hole masses) and frequency ({\em i.e.,} orbital
period) when the time-dependent perturbation can have a significant
effect. A second involves determining the degree to which the efficacy of
energy and mass transport reflect the degree of chaos exhibited by the orbits,
both in the presence and the absence of the perturbation. To
what extent, {\em e.g.,} does efficient energy transport require that
a large fraction of the orbits in the time-dependent potential be chaotic?
Does resonant phase mixing rely crucially on the presence of transient chaos?

Another issue involves determining the extent to which
the bulk manifestations of a black hole binary vary for spherical,
axisymmetric, and nonaxisymmetric ({\em e.g.,} triaxial) galaxies. Is it,
{\em e.g.,} true that spherical and nearly spherical systems are impacted
less by the presence of a supermassive binary since, in the absence of the
binary, all or almost all of the orbits are regular?
In a similar vein, one would like to understand the extent to which
the effects of the binary depend on the steepness of the cusp. And, perhaps
most importantly, it would seem crucial to determine how the size of the
`sphere of influence' of the binary depends on the size of the black hole
orbits and their masses. Perhaps the most important conclusion here
is that this `sphere'  can be much larger than the
size of the black hole orbits. For plausible choices of parameter values,
black holes moving along orbits with size ${\sim}{\;}r_{h}$ can significantly
impact the density distribution at radii as large as ${\sim}{\;}10-20r_{h}$
or more.

All these issues have important implications for determining when a 
supermassive black hole binary might be expected to have observable 
consequences. The second half of the paper, Sections 3 and 4, considers these 
consequences. Section 3 considers the generality of the simple
models considered in Section 2, which assume circular orbits and equal
mass black holes, and then focuses on
direction-dependent effects which must be understood to determine
how potentially observable quantities depend on the relative orientation of
the observer and the binary.

Section 4 focuses in detail on one specific observable prediction,
namely that supermassive black hole binaries can alter the density
distribution near the center of a galaxy. This involved: (i) generating
$N$-body realisations of density distributions consistent with a Nuker Law
(Lauer {\em et al} 1995); (ii) evolving these $N$-body systems in the fixed
time-dependent potential corresponding to the galaxy plus orbiting black holes;
(iii) determining how the initial density distribution changes over the
course of time; and, (iv) presuming that mass traces light, integrating the
resulting density distribution along the line of sight to obtain a surface
brightness profile. These are {\em not} self-consistent computations;
but they {\em can} at least provide strong indications as to what the expected
effects of the binary would be. The crucial point, then, is that such an
exercise results generically in brightness distributions that resemble
qualitatively the forms reported in Lauer {\em et al} (2002); and that by
fine-tuning parameters within a reasonable range, one can reproduce many of
the details of what is actually observed.

Section 5 summarises the principal conclusions and discusses potential
implications.

\section{Dynamical Effects of Supermassive Black Hole Binaries}

\subsection{Description of the experiments}
The computations described here involved orbits evolved in potentials of
the form
\begin{equation}
V(x,y,z)=V_{0}(x,y,z)
-{M\over |{\bf r}-{\bf r}_{1}(t)|}
-{M\over |{\bf r}-{\bf r}_{2}(t)|},
\end{equation}
where $V_{0}$ is time-independent and 
${\bf r}_{1}$ and ${\bf r}_{2}$ correspond to circular orbits in
the $x-y$ plane, {\em i.e.,}
\begin{equation}
x_{1}(t)=r_{h}\sin {\omega}t , \qquad y_{1}(t)=r_{h}\cos {\omega}t, \qquad
z_{1}(t)=0,
\end{equation}
and ${\bf r}_{2}=-{\bf r}_{1}.$
Some of the computations focused on a harmonic oscillator potential,
\begin{equation}
V_{0}(x,y,z)={1\over 2}m^{2},
\end{equation}
with
\begin{equation}
m^{2}=(x/a)^{2}+(y/b)^{2}+(z/c)^{2}.
\end{equation}
Others focused on more realistic potentials of the form
\begin{equation}
V_{0}(x,y,z)=-{1\over (2-{\gamma})}
\left[1- {m^{2-{\gamma}}\over (1+m)^{2-{\gamma}}} \right],
\end{equation}
with ${\gamma}$ is a cusp index, assumed to satisfy $0{\;}{\le}{\;}{\gamma}
{\;}{\le}{\;}2$.
The axis ratios $a$, $b$, and $c$ were selected of order unity.

The assumptions that the black holes are in circular orbits and
that they have equal masses might appear an extreme idealisation.
However, as will be
discussed in Section 3, it appears that relaxing these assumptions does
{\em not} change the principal conclusions. {\em This model
appears structurally stable towards modest changes in the orbital parameters
of the binary.}

For $a=b=c=1$, eq.~(5) reduces to the spherical Dehnen (1994) potential 
with unit mass, and, quite generally, for large $r$, $V \to -1/m$. Thus
for axis ratios of order unity, one can interpret
eq.~(5) as the potential for a galaxy with mass $M_{g}{\;}{\approx}{\;}1.0$.
For nonspherical systems, eq.~(5) yields density distributions different
from Merritt and Fridman's (1996) triaxial Dehnen models,
in that it is $V$, rather than ${\rho}$, that is constrained to
manifest ellipsoidal symmetry.

This potential is unrealistic in that,
for large radii, $V$ does not become spherically symmetric; and one can also
argue that it is unrealistic in the sense that, assuming mass traces light,
the isophotes become peanuty for axis ratios far from spherical. Given,
however, that one is interested primarily in physical processes in the
central portions of the galaxy, the $r\to\infty$ asymptotic behaviour is
largely unimportant; and it should be recalled that the isophotes in `real'
galaxies tend to manifest systematic deviations from ellipticity ({\em e.g.,}
Kormendy \& Bender 1996). This potential has the {\em huge} advantage that,
unlike Merritt and Fridman's nonspherical Dehnen potential, it can be
expressed analytically, thus reducing by two orders of magnitude or more
the time required for orbital integrations. Moreover, as discussed in Section
4, for the case of spherical symmetry the behaviour of orbits in this
potential is very similar to orbits evolved in the potential associated with
a Nuker Law, at least for those choices of Nuker parameters for which the
potential can be expressed analytically.

For spherical systems with $a=b=c=1$, 
\begin{equation}
M(r)=\left[ r/(1+r)\right]^{(3-{\gamma})}
\end{equation}
is the mass within $r$ of the galactic center.
For axis ratios of order unity, eq.~(7) also provides a
reasonable estimate for moderately nonspherical systems.

The computations here involved black hole masses in the range 
$0.005{\;}{\le}{\;}M{\;}{\le}{\;}0.05$ and radii satisfying 
$0.005{\;}{\le}$ $r_{h}{\;}{\le}{\;}0.5$. 
Following Merritt and Fridman's (1996) normalisation for
their triaxial Dehnen model, one can translate the
dimensionless model into physical units by defining the correspondence
\begin{equation}
t=1 \qquad \Leftrightarrow \qquad
1.46\times 10^{6} M_{11}^{-1/2}a_{\rm kpc}^{3/2}\,{\rm yr},
\end{equation}
where
\begin{displaymath}
M_{11}= \left( {M\over 10^{11}M_{\odot}} \right) \qquad {\rm and} \qquad
a_{\rm kpc}= \left( {a\over 1\, {\rm kpc} } \right).
\end{displaymath}

One can identify an energy-dependent dynamical time $t_{D}$ 
following either Merritt and Fridman, who related it to the period of a
specific type of regular orbit, or Kandrup \& Siopis (2003), who
proposed a prescription based on the times between turning
points in representative orbits. Those two prescriptions yield results in
agreement at the 5\% level or better.
More generally, for axis ratios of order unity, at least for small radii
the angle-averaged density and energy distributions are relatively
similar to those associated with `true' maximally triaxial Dehnen models,
so that a dynamical time $t_{D}(E)$ can be estimated to within 20\% or so
from Table 1 in Merritt \& Fridman (1996). This implies, {\em e.g.,} that,
for ${\gamma}=1.0$, a time $t=512$ corresponds to roughly $100t_{D}$ for
stars in the 20\% mass shell or, equivalently, ${\sim}{\;}8\times 10^{8}
M_{11}^{-1/2}a_{\rm kpc}^{3/2}$ yr.

A `realistic' value for the frequency 
can be estimated easily.
Suppose, {\em e.g.,} that $a=b=c=1.0$. If $M{\;}{\ll}{\;}M(r_{h})$, with
$M(r_{h})$ the galactic mass contained within radius $r_{h}$, the
black holes can be viewed as test particles moving in the galactic potential.
This implies that 
\begin{equation}
{\omega}^{2}=r_{h}^{-{\gamma}}(1+r_{h})^{{\gamma}-3},
\end{equation}
so that, {\em e.g.,}
${\omega}=r_{h}^{-1/2}(1+r_{h})^{-1}$ for ${\gamma}=1.0$. If, alternatively, 
$M{\;}{\gg}{\;}M(r_{h})$ the potential associated with the galaxy can be
neglected and one is reduced {\em de facto} to the circular, equal mass
two-body problem, for which ${\omega}=\sqrt{M/4r_{h}^{3}}$. 
For ${\gamma}=1.0$, $M=0.01$, and $r_{h}=0.05$, fiducial values considered
in many of the computations, the galactic potential can be neglected in
a first approximation, so that ${\omega}{\;}{\approx}{\;}\sqrt{20}{\;}{\approx}
{\;}4.47.$ 

However, for much of this section, ${\omega}$ was viewed as
a free parameter so that, for fixed amplitude and geometry, one can explore
the response as a function of driving frequency. This enables one to determine
the extent to which the response manifests a sensitive dependence on frequency,
which can provide important insights into the resonant couplings 
generating the response.

Attention focused primarily on the statistical properties of representative
orbit ensembles, integrated from sets of ${\ge}{\;}1600$ initial conditions.
These were generated by uniformly sampling a specified constant energy
hypersurface
{\em as defined in the limit $M\to 0$} using an algorithm described in Kandrup
\& Siopis (2003). Allowing for the black holes
changes the initial energies, so that one is {\em de facto} sampling a
`slightly thickened' constant energy hypersurface.
The initial conditions were integrated forward for a time $t=512$. 
The integrations also tracked the evolution
of a small initial perturbation, periodically
renormalised in the usual fashion ({\em e.g.,}
Lichtenberg \& Lieberman 1992), so as to extract estimates of the largest
finite time Lyapunov exponent. 

For each simulation, specified by $a$, $b$, $c$, $M$, $r_{h}$, and ${\omega}$,
following quantities were extracted:
\par\noindent (i) the fraction $f$ of `strongly chaotic' orbits, estimated
as in Kandrup \& Siopis (2003) (as discussed in Kandrup, Vass \& Sideris
[2003], because the potential is time-dependent it is often difficult
to make an absolute distinction between regular and chaotic orbits, although
it {\em is} relatively easy to identify orbits that are `strongly chaotic');
\par\noindent (ii) the mean value ${\langle}{\chi}{\rangle}$ of the finite
time Lyapunov exponents for the strongly chaotic orbits;
\par\noindent (iii) the mean value ${\langle}{\delta}E{\rangle}$ of the
energy shift ${\delta}E{\;}{\equiv}{\;}E(t)$ $-E(0)$ for all the orbits at
various times $t>0$;
and
\par\noindent (iv) the dispersion ${\sigma}_{{\delta}E}$ associated with
these shifts.
\par\noindent
The data were also analysed to determine the functional forms of
${\langle}{\delta}E(t){\rangle}$ and ${\sigma}_{{\delta}E}(t)$, and to search
for correlations between changes in energy and values of finite time Lyapunov
exponents for individual orbits within a single ensemble.

Other integrations tracked phase mixing in initially localised ensembles, 
so as to determine the extent to which  such mixing 
resembles ordinary chaotic phase mixing in a time-independent potential
({\em e.g.,} Merritt \& Valluri 1996, Kandrup 1998) or resonant phase mixing
in a galaxy subjected to large scale bulk oscillations (Kandrup, Vass \&
Sideris 2003).

\subsection{Statistical properties of orbit ensembles}
Overall, as probed by the shuffling of orbital energies, there is a
broad and comparatively efficient resonant response. For fixed values
of $a$, $b$, $c$, $M$, and $r_{h}$, the range of `interesting' frequencies
${\omega}$ can be two orders of magnitude or
more in breadth. {\em One does not need to `fine-tune'
${\omega}$ to trigger an efficient shuffling of energies.} However, the 
resonance {\em can} exhibit substantial structure, especially for the case of 
a spherically symmetric oscillator potential. Superimposed upon a smooth 
overall trend, quantities like ${\langle}{\delta}E{\rangle}$ can exhibit a 
complex, rapidly varying dependence on ${\omega}$.

Consider, {\em e.g.,} Fig.~1, which plots ${\langle}{\delta}E{\rangle}$ and 
${\sigma}_{{\delta}E}$ as functions of ${\omega}$ at time $t=512$ for two
oscillator models, one spherical and the other triaxial. Both have $M=0.05$ 
and $r_{h}=0.3$ The curves for the two models have envelopes with a 
comparatively simple shape but, for the spherical model, an enormous amount 
of substructure is superimposed. This substructure reflects the fact
that all unperturbed 
orbits oscillate with the same frequency. Indeed, close examination
reveals that the resonances are associated
with integer and (to a lesser degree) half-integer values of ${\omega}$,
harmonics of the unperturbed natural frequency ${\omega}=1.0$.
In an axisymmetric system, there are two natural frequencies,
which can yield a yet more complex response pattern. If, however, $M$ and
$r_{h}$ are chosen large enough to elicit a significant response,
the resonances typically broaden to the extent that much, if not all, that
structure is lost. Allowing for a fully triaxial system leads to three
unequal frequencies, which yields such a plethora of harmonics
that, even for comparatively weak responses, the short scale structure is
largely lost.

It is evident from Fig.~1 that, although ${\langle}{\delta}E{\rangle}$ is
substantially larger for the triaxial than for the spherical model,
${\sigma}_{{\delta}E}$ is comparable. What this means is that, even though
the spherical model leads to a smaller systematic shifting in energies, the
energies of orbits in these two models are shuffled to a comparable degree.
The observed differences in
${\langle}{\delta}E{\rangle}$ do {\em not} reflect the fact that the
nonspherical model is triaxial. Rather, they appear again to reflect the
fact that, for a spherical system, there is only one characteristic frequency
for the unperturbed orbits. Modest deviations from spherical symmetry, be
these either axisymmetric or not, suffice typically to yield amplitudes
more closely resembling Fig.~1 (c) than 1 (d).

As indicated in Fig.~2, there is 
a threshold value of $M$ below 
which no substantial response is observed; and, similarly,
the response `turns off' for higher-energy orbits that spend most of their
time far from the binary. However, for `interesting' choices of $M$, as probed,
{\em e.g.,} by ${\langle}{\delta}E({\omega}){\rangle}$ or
${\sigma}_{{\delta}E}({\omega})$, the resonance has a characteristic shape.
As the frequency increases from ${\omega}=0$, ${\langle}{\delta}E{\rangle}$
and  ${\sigma}_{{\delta}E}$ exhibit a rapid initial increase, peak at a
maximum value, and then begin a much slower decrease.
For fixed parameter values, {\em the value of the frequency
triggering the largest response is roughly
independent of mass,} but it {\em is} true that, for larger $M$,
the relative decrease in  ${\langle}{\delta}E({\omega}){\rangle}$ and
${\sigma}_{{\delta}E}({\omega})$ with increasing ${\omega}$ is slower than
for smaller $M$. This is consistent with the notion that, for larger-amplitude 
perturbations, higher-order harmonics become
progressively more important.  Fig.~2 also shows the peak frequency is a 
decreasing function of $r_{h}$. In particular, when the black holes are closer 
together one requires larger ${\omega}$ 
to elicit a significant response.

Efficient shuffling of energies seems tied unambiguously to the presence of
large amounts of chaos, as probed by the fraction $f$ of strongly chaotic
orbits and, especially, the size of a typical finite Lyapunov exponent ${\langle}{\chi}{\rangle}$. Large $f$ and ${\langle}{\chi}{\rangle}$ do not guarantee large changes
in energies, but they are an essential prerequisite. In some cases, notably
nearly spherical systems, $f$ and ${\langle}{\chi}{\rangle}$ are
very small in the limit ${\omega}\to 0$. As ${\omega}$ increases, however,
$f$ and especially ${\langle}{\chi}{\rangle}$ also increase; and, for
values of ${\omega}$ sufficiently large to trigger an efficient response,
the ensemble will be very chaotic overall. For values of ${\omega}$ in the
resonant region, $f$ and ${\langle}{\chi}{\rangle}$
tend to exhibit only a comparatively weak dependence on ${\omega}$. 

Orbit ensembles evolved in spherical, axisymmetric, and triaxial 
Dehnenesque potentials exhibit resonance patterns quite similar
both to one another and to the patterns observed in nonspherical oscillator
models, although some relatively minor differences {\em do} exist. 
Fig.~3 exhibits data for two such models, one spherical and the 
other triaxial, each with ${\gamma}=1.0$, $r_{h}=0.05$, and $M=0.01$. 
The models were both generated for ensembles of initial conditions with
$E=-0.70$ and ${\langle}r_{in}{\rangle}{\;}{\approx}{\;}0.33$.
The black hole radius $r_{h}=0.05$ corresponds roughly to the 0.2\% mass shell.

The triaxial oscillator model in Fig.~1 and the Dehnenesque models in Fig.~3
are representative in that modest changes in the parameters of the binary 
do not lead to significant qualitative changes in the response. Equally 
important, however, they are also robust towards changes in axis ratio.
Axisymmetric and slightly triaxial Dehnenesque models ({\em e.g.,} as 
nonspherical as $a^{2}=1.05$, $b^{2}=1.00$, and $c^{2}=0.95$)
yield results very similar to the spherical case. Other `strongly' triaxial 
models yield results similar to the particular triaxial model exhibited here.

Even for spherical and nearly spherical systems, the relative measure 
$f$ of strongly chaotic orbits tends to be large even for ${\omega}=0$, this 
corresponding to stationary but separated black holes.
{\em One does not require a strong time-dependence to generate a
large measure of chaotic orbits}. However, it {\em is} true that, for
axisymmetric and other nearly spherical systems, the size of a typical Lyapunov
exponent tends to be considerably smaller than for strongly triaxial systems.

In significantly triaxial models, the degree of chaos, as probed by $f$ or
${\langle}{\chi}{\rangle}$, is a comparatively flat function of ${\omega}$.
By contrast, in nearly spherical and axisymmetric systems, the degree of chaos,
especially as probed by ${\langle}{\chi}{\rangle}$, increases rapidly with
increasing ${\omega}$ until it becomes comparable to the degree of chaos
exhibited by strongly triaxial models. The obvious inference is that, 
{\em when the system is nearly spherical or axisymmetric, the time-dependence
associated with the orbiting binary is required to give the chaotic orbits
particularly large Lyapunov exponents.} 

An analogous result 
holds for the mean energy shift
${\langle}{\delta}E{\rangle}$. Quite generally,
${\langle}{\delta}E{\rangle}\to 0$ for ${\omega}\to 0$ and increases with 
increasing ${\omega}$. However, the initial rate of increase is typically 
much larger for significantly triaxial models than for nearly spherical and 
axisymmetric systems. This has an important practical implication: 
Because larger frequencies are required
to trigger the resonance in galaxies that are nearly axisymmetric, 
{\em in nearly spherical or axisymmetric galaxies black holes of given
mass must be in a tighter orbit before they can trigger a significant 
response.} 

\subsection{Shuffling of energies as a diffusion process}
Overall, the shuffling of energies induced by the black hole binary is
diffusive, although the basic picture depends on the amplitude of
the response. When changes in energy experienced by
individual orbits are relatively small, the dispersion tends
to grow diffusively, {\em i.e.,} 
${\sigma}_{{\delta}E}{\;}{\propto}{\;}t^{1/2}.$
The mean shift in energy typically grows more quickly,
being reasonably well fit by a linear growth law
${\langle}{\delta}E{\rangle}{\;}{\propto}{\;}t$.
Alternatively, when the response is stronger, it is the mean shift that
grows diffusively, {\em i.e.,} 
${\langle}{\delta}E{\rangle}{\;}{\propto}{\;}t^{1/2},$ whereas the dispersion 
is well fit by a growth law
${\sigma}_{{\delta}E}{\;}{\propto}{\;}t^{1/4}$. Examples of both sorts of
behaviour can be seen in Fig.~4.

One might 
have supposed that, since the shuffling of energies is
associated with the presence of chaos, changes in energy should grow
exponentially. This however, does {\em not} appear to be the case, at least
macroscopically. The initial response of the orbits
($t<5\,t_{D}$ or so) may be exponential, but it is evident that, overall, the
response is diffusive. {\em Time-dependent chaos does {\em not} trigger
exponentially fast mixing in energies. However, it can still be extremely
important
in that it allows comparatively efficient shufflings of energies that would be
completely impossible in a time-independent Hamiltonian system.}

One final point should be stressed. That changes in energy are diffusive,
reflecting a slow accumulation of energy shifts, corroborates a
fact also evident from an examination of individual orbits:
Changes in energy experienced by individual orbits do {\em not} result from
single close encounters with the black holes. Instead, they really do reflect
resonance effects associated with the time-dependent potential.

\subsection{Correlations amongst orbital properties for different
orbits within an ensemble}
Orbits with smaller finite time Lyapunov exponents ${\chi}$ tend to exhibit
energy shifts that are smaller in magnitude $|{\delta}E|$. Orbits with large
${\chi}$ can experience both large and small net changes in energy.
As has been observed in other time-dependent potentials (Kandrup \&
Terzi{\'c} 2003), the fact that an orbit is chaotic does not
necessarily imply that it will exhibit large, systematic drifts in energy 
over a finite time interval. However, the energy shifts in orbits with 
small ${\chi}$ are invariably small.

When the response is weak, so that the dispersion of the ensemble 
evolves diffusively, changes in energy exhibited by individual orbits
are comparably likely to be positive or negative. However, when the response
is stronger, so that the mean shift evolves diffusively, the 
energies of individual orbits tend instead to increase systematically.

When the response is relatively weak and changes in energy are equally likely
to be either positive or negative, the distribution of energy shifts
$n({\delta}E)$ is typically well fit by a Gaussian with mean roughly equal
to zero. However, when the response becomes stronger, $n({\delta}E)$
becomes distinctly asymmetric and cannot be well fit by a
Gaussian, even allowing for a nonzero mean.

Correlations between the `degree' of chaos and the `degree' of energy shuffling
experienced by individual orbits are perhaps best illustrated by extracting
energy shifts ${\delta}E$ at different times $t_{i}$ {\em for individual
orbits}, computing the mean and dispersion, ${\langle}{\delta}E{\rangle}$
and ${\sigma}_{{\delta}E}$, associated with the resulting time series
$\{ {\delta}E(t_{i}) \}$, and demonstrating how these moments correlate with
the value of the finite time Lyapunov exponent ${\chi}$. Examples of such an
analysis are exhibited in Fig.~5. The obvious point is
that the moments are invariably small when ${\chi}$ is small, whereas larger
${\chi}$ typically implies larger values of $|{\langle}{\delta}E{\rangle}|$
and ${\sigma}_{{\delta}E}$.

\subsection{Chaotic and resonant phase mixing}
The time-dependent potential associated with the black hole binary can
alter `ordinary' chaotic phase mixing in at least two important ways.

The time-dependent potential tends to increase both the fraction of chaotic 
orbits and the size of a typical Lyapunov exponent. If a galaxy is in a 
(nearly) time-independent (near-)equilibrium state,
the relative measure of (at least strongly) chaotic orbits should be
relatively small, since presumably one requires large measures of regular
(or nearly regular) orbits to provide the `skeleton' of the interesting
structures associated with those chaotic orbits which {\em are}
present (Binney 1978). Introducing a time-dependent perturbation
leads oftentimes to a significant increase in the relative measure of chaotic
orbits. Moreover, even when the time-dependence does not significantly
increase the measure of chaotic orbits, it can make already chaotic orbits
more unstable, thus allowing them to mix more efficiently.

Because energy is no longer conserved, the time-dependent potential also
allows mixing between different constant energy hypersurfaces, which 
is completely impossible in the absence of a time-dependence.
Overall, this mixing of energies is not as
efficient a process as mixing in configuration or velocity space. However, the
resonant mixing of energies associated with chaotic orbits still plays an
important role. 

An example of such resonant phase mixing is provided in the left panels
of Fig.~6, which track an initially localised ensemble with
$E=-0.70$ in a spherical Dehnen potential with
${\gamma}=1$ and $a=b=c=1.0$, allowing for black hole parameters $M=0.005$,
$r_{h}=0.05$, and ${\omega}=\sqrt{10}$. The right panels track the same
ensemble, evolved identically except that ${\omega}=0$.
Two points are immediate. One is that, for the realistic
case when ${\omega}{\;}{\ne}{\;}0$, a time $t=64$, corresponding to
${\sim}{\;}10^{8}M_{11}^{-1/2}a_{\rm kpc}^{3/2}$ yr, is sufficient to achieve
a comparatively well mixed
configuration. Achieving a comparable degree of mixing for the ${\omega}=0$
system requires a time $t>512$. The other point is that orbits in the
ensemble evolved with ${\omega}{\;}{\ne}{\;}0$ have diffused to radii
$r>0.3$, which is impossible for orbits in the ${\omega}=0$
ensemble, for which energy is conserved.

\section{Observational Consequences of the Dynamics}
\subsection{Generality of the idealised model}
\vskip .1in
Attention hitherto has focused on the dynamical consequences of
a supermassive black hole binary, viewed as the prototype of a time-dependent
perturbation acting in a galaxy idealised
otherwise as a collisionless
equilibrium. The object of this and the following section is to consider
instead potentially observable consequences, the most obvious of which is
a changing surface brightness distribution induced by a readjustment in the
mass density as stars are transported to larger radii.

In so doing, one can proceed
by viewing the host galaxy as a superposition of orbit ensembles with
different energies $E$ and, for various choices of binary parameters,
determining when, for any given value of $E$, the binary can have an 
appreciable effect, {\em
e.g.,} by generating a large energy shift ${\langle}{\delta}E{\rangle}$. 
As described already, the response will
only be large when the size $r_{h}$ of the binary orbit is sufficiently small
that the total black hole mass $M_{1}+M_{2}{\;}{\ge}{\;}M(r_{h})$. This,
however, implies that, in a first approximation, 
the frequency of the binary can be estimated neglecting the bulk potential
of the galaxy. Thus, relaxing the assumptions of equal masses
and strictly circular orbits,
\begin{equation}
{\omega}= \sqrt{M_{1}+M_{2}\over A^{3}},
\end{equation}
with $A$ the value of the semi-major axis.

Perhaps the most obvious question here is simply: For fixed $E$ and $A$,
how do quantities like ${\langle}{\delta}E{\rangle}$ depend on the total
mass $M_{tot}=M_{1}+M_{2}$? The answer is that, at least for `realistic'
binary black hole masses, {\em i.e.,} $M_{1}$ and
$M_{2}<0.01M_{gal}$, ${\langle}{\delta}E{\rangle}$ is a
comparatively smooth, monotonically increasing function of $M_{tot}$.
For very small masses, there is essentially no response; but, beyond a critical
mass, the precise value of which depends on properties of the host galaxy,
the dependence is roughly power law in form, {\em i.e.,}
${\langle}{\delta}E{\rangle}{\;}{\propto}{\;}M_{tot}^{p}$, with
the power $p$ typically in the range $1<p<2$.
Examples of this behaviour are exhibited in the left panels of Fig.~7,
which show the effects of increasing the total mass for five different models,
one spherical, one prolate axisymmetric, one oblate axisymmetric, and two
genuinely triaxial. This particular set of examples again incorporated circular
orbits and equal black hole masses; but, as will
be discussed below, these assumptions are not crucial.

A second obvious question is: How small must the binary orbit be in order to
elicit a significant response? Physically, one might suppose that the binary
was initialised in a comparatively large orbit as the result of a merger of two
colliding galaxies; but that the orbit slowly decayed via dynamical friction,
allowing the black holes to sink toward the center of the galaxy. However,
within the context of such a scenario the crucial issues to determine are (i)
when the binary can begin to have a large effect, {\em i.e.,} how
small the orbit must be; and (ii) when the effects of the binary `turn off'
again. These issues are addressed in the right panels of Fig.~7, which
exhibit ${\langle}{\delta}E{\rangle}$ as a function of $r_{h}$ for the same
five galactic models used to generate the left panels.

Two points are evident: (1) The binary has its largest effect when
$r_{h}$ is substantially smaller than the typical radius of the orbits with
the specified energy. The ensembles considered were each comprised of
orbits with initial energy $E=-0.70$ and mean radius
${\langle}r{\rangle}{\;}{\approx}{\;}0.33$, but the maximum response was
observed for $r_{h}{\;}{\sim}{\;}0.04$, {\em i.e.,} a size
roughly ten times smaller! This 
reflects the fact that mass and energy transport have been triggered by a
resonance, rather than by direct binary scatterings of individual stars with
the black holes. One needs a very tight binary orbit to get frequencies
sufficiently large to trigger a significant response. (2) As noted already,
for the triaxial models the effects of the binary 
`turn on' at substantally larger values of $r_{h}$ than for the spherical
and axisymmetric systems. This would suggest that a black hole binary could
have an especially large effect in a strongly triaxial galaxy: since the
range of black hole sizes that can have an appreciable effect is
substantially larger, the time during which the resonance will act should
presumably be longer.

But how generic are the idealised computations described in Section 3?
It might not seem unreasonable to assume that the black holes follow nearly
circular orbits, since dynamical friction will tend to circularise
initially eccentric orbits; but the assumption of equal mass black holes is
clearly suspect.

Computations show that varying the eccentricity $e$ within reasonable bounds 
has only a comparatively minimal effect. Increasing $e$ from
values near zero to a value as large as $e=0.5$ will not change quantities
like ${\langle}{\delta}E{\rangle}$ by more than 25\%; and, in general the
effect is much smaller even than this. This is, {\em e.g.,} evident from
the left panels of Fig.~8, which were generated for the same five
models considered in Fig.~7.

As is evident from the right hand panels of Fig.~8, 
there is a substantially stronger, systematic dependence on the mass ratio
$M_{2}/M_{1}$. For fixed $M_{tot}=M_{1}+M_{2}$, the largest effects arise for
$M_{1}{\;}{\approx}{\;}M_{2}$; but even here the dependence on the mass
ratio is not all that sensitive. In particular, for all but the triaxial
models, the response is a relatively flat function of $M_{2}/M_{tot}$
for $M_{2}/M_{tot}{\ga}0.25$, so that, for fixed $M_{1}+M_{2}$, mass ratios
$1/3 {\;}{\la}{\;}M_{2}/M_{1}{\;}{\la}{\;}1$ yield comparable results.
It {\em is} true that, for fixed semi-major axis $A$ and total mass
$M_{tot}$, the effect of the binary is significantly reduced for
$M_{2}{\;}{\ll}{\;}M_{1}$, but the reason for this is obvious:
When $M_{2}{\;}{\ll}{\;}M_{1}$, the more massive black hole is located very
near the center of the galaxy. This implies, however, that, even if the binary
has a very high frequency, the more massive black hole remains too close to
the center to have an appreciable effect 
at large radii.
The smaller black hole is typically found at much larger values of $r$, but
its mass is too small to have a significant effect.

\subsection{Systematic changes in density}
Changes in energy induced by transient chaos lead generically to a
readjustment in bulk properties like density; and, to the extent that there
is an average increase in energy, this readjustment implies a systematic
displacement of stars to larger radii. To see how this effect can proceed,
one can sample a constant energy hypersurface to generate a set of initial
conditions, integrate those initial conditions into the future, and then
compare angle-averaged radial density distributions ${\rho}(r)$ generated at
various times $t{\;}{\ge}{\;}0$.

The left panels of Fig.~9 summarise results for a 
model with $a^{2}=1.25$, $b^{2}=1.0$, and
$c^{2}=0.75$, assuming circular orbits with $M_{1}=M_{2}=0.01$, 
$r_{h}=0.05$, and
${\omega}=\sqrt{20}$. The ensemble 
was so constructed that $E=-0.70$ and
${\langle}r_{in}{\rangle}{\;}{\approx}{\;}0.33$. The five panels exhibit
the density distributions at $t=0$, $16$, $32$, $64$, and $128$, the
last corresponding physically to 
${\sim}{\;}2\times 10^{8}M_{11}^{-1/2}a_{\rm kpc}^{3/2}$ yr.
The right panels exhibit analogous data for the same ensemble 
and potential but with the black holes held fixed in space,
{\em i.e.,} ${\omega}{\;}{\equiv}{\;}0$.

The density distribution remains
essentially unchanged for the time-independent ${\omega}=0$ potential,
but the realistic case with ${\omega}=\sqrt{20}$ leads to a significant 
density readjustment. (Minor changes in the ${\omega}=0$
model reflect a modest readjustment to the insertion of the fixed binary in
an equilibrium generated without a binary.) Already by $t=16$ (11 binary periods), corresponding to an interval ${\sim}{\;}2.5\times 10^{7}M_{11}^{-1/2}a_{\rm kpc}^{3/2}$ yr, there is a pronounced decrease in density in the range $0.3{\;}{\le}{\;}r{\;}{\le}{\;}0.5$  and an increase in density at larger radii. Initially the trajectories are restricted energetically to $r{\;}{\le}{\;}0.6$. By $t=128$, more than 13\% of the trajectories are
located at $r>1.0$.

\subsection{The size of the `sphere of influence'}
Figure 9 demonstrates that a black hole binary can significantly
impact orbits which spend most of their times at radii 
${\gg}{\;}r_{h}$. The obvious question, however, is: how much larger? 
To answer this question one can evolve ensembles with a variety of different
initial radii, and determine their response as a function of $r$.
The results of two such investigations are summarised in Fig.~10. In each
case, the configuration corresponded to a spherical Dehnen model with
$a=b=c=1.0$ and a 
binary with $M_{1}=M_{2}=0.005$, 
$r_{h}=0.25$, and ${\omega}=0.2828$. The left  panels are for a model with
${\gamma}=0.0$; the right  panels for ${\gamma}=1.0$.

The `sphere of influence' is in fact quite large, extending 
out to $r{\;}{\ge}{\;}4$, even though $r_{h}=0.25$. Moreover, it is evident
that the ensembles which experience the most shuffling in energies,
as probed by ${\langle}|{\delta}E|{\rangle}$ and ${\sigma}_{{\delta}E}$, are
precisely those ensembles with the largest Lyapunov exponent
${\langle}{\chi}{\rangle}$. Indeed, for the ${\gamma}=0.0$ and
${\gamma}=1.0$ models, the rank correlation between the mean shift
${\langle}|{\delta}E|{\rangle}$ and the mean exponent
${\langle}{\chi}{\rangle}$
for different ensembles, are, respectively,
${\cal R}({\langle}{\delta}E{\rangle},{\langle}{\chi}{\rangle})
=0.615$ and $0.613$.

It is also clear that the value of the cusp index ${\gamma}$ has a
significant effect on the details of the response. The value
of ${\gamma}$ does {\em not} have a large effect on the size of the binary
`sphere of influence', but it {\em does} impact the amplitude of the response
and how that response correlates with radius. In both cases, there is a
significant response for $0.15{\;}{\le}{\;}r{\;}{\le}{\;}6.0$, but the
response in this range, as probed both by the degree of shuffling in energies,
is somewhat larger for the cuspy model. Even more strikingly, however, 
the cusp appears to reduce both the size of the
Lyapunov exponents and the degree of shuffling at very small radii. For the
cuspy model with ${\gamma}=1.0$, comparatively little shuffling of energies
and comparatively small amounts of chaos are observed at radii
${\ll}{\;}r_{h}$. The very lowest energy stars tend to be
more regular and to be less susceptible to resonant mixing.

\subsection{Anisotropy}
To what extent does the mass transport induced by a supermassive black hole
binary depend on direction? Even if, {\em e.g.,} the host galaxy is modeled
as exactly spherical, the binary breaks the symmetry and, as such, could
introduce anisotropies into a completely isotropic ensemble of stars. This
is important since such anisotropies would imply that changes in visual 
appearance induced by the binary could depend appreciably on the observer's
viewing angle.

As a simple example, one can consider the direction-dependent density 
distributions associated with a uniform sampling of a constant energy
hypersurface which, assuming a spherical potential,
implies a spherically symmetric density distribution and an isotropic
distribution of velocities. One example thereof is exhibited in Fig.~11,
which was generated for a ${\gamma}=1.0$ Dehnen model with $a=b=c=1$ 
containing a binary executing a circular orbit in the $x-y$ plane with
the `correct' Kepler frequency. Here the top left panel exhibits spatial
distributions 
at times $t=0$ and $t=512$; the top right 
panel shows the corresponding velocity distributions.
At $t=0$, the spatial and velocity
distributions are equal modulo statistical uncertainties; at late times
they differ systematically, but it remains true that $n(|x|)=n(|y|)$ and
$n(|v_{x}|)=n(|v_{y}|)$. There is clearly 
a systematic outward transport of stars in all three directions,
but it is also evident that there is a
larger net effect on the spatial components in the plane of the orbit. 
Similarly, there is a modest shift in velocities which, again, is more
pronounced in the $x$ and $y$ components. 
The bottom two panels contain plots of, respectively, the $x$ and $y$ and
the $x$ and $z$ coordinates at $t=512$.
These panels confirm that the final 
distribution is more extended in the plane of the binary than in the
orthogonal direction. In particular, it 
could easily be misinterpreted as a disc or a torus. 

But what if the host galaxy is already nonspherical? If, {\em e.g.,} 
the galaxy is genuinely triaxial, one might 
suppose that the binary 
will have settled into a symmetry plane; but, assuming that this be
the case, there are at least two obvious questions.
(1) How does the overall response depend on which symmetry plane?
(2) For a binary oriented in a given plane,
to what extent do observable properties depend on viewing angles?

Both these questions were addressed as before by evolving uniform samplings
of constant energy hypersurfaces which yield a triaxial number density 
but are still characterised by an isotropic distribution of velocities. 
The result of one such computation is summarised in Fig.~12. Here panel (a)
exhibits the initial density distributions; 
the remaining
three panels exhibit the corresponding distributions at $t=512$ for three
different integrations, with the binary oriented in the $x-y$, $y-z$, and
$z-x$ planes. 

Overall the `angle-averaged' properties of the different
simulations are very similar: The mean short time Lyapunov exponents 
${\langle}{\chi}{\rangle}$ for the three different runs agree to within
$10\%$, and even smaller variations were observed for quantities like 
${\langle}{\delta}E{\rangle}$. Indeed, the shape of the galaxy seems more 
important than the orientation of the binary. For all three binary 
orientations, one observes that the largest effect is in the $x$-direction, 
which corresponds to the long axis, and the smallest in the short-axis
$z$-direction. The details of the response observed here depend to a
considerable extent on both the shape of the potential and the energy of
the initial ensemble. In particular, for some choices the response is 
largest in the short-axis rather than long-axis direction. However, it seems 
true quite generally that the orientation of the binary is comparatively 
unimportant. There remains a dependence on viewing angle but, if anything, 
this effect is somewhat weaker than for the case of spherical systems. 

For axisymmetric systems with the binary oriented in the
$x-y$ symmetry plane, one finds generically that distributions in the $x$
and $y$ directions agree to within statistical uncertainties, but that 
the $z$-direction distributions differ systematically. In some cases
(depending on both shape and binary parameters),
there is more mass transport in the $z$ direction; in others the effect
is more pronounced in the $x$ and $y$ directions. These differences likely
reflect the fact that this mass transport is triggered by a resonance. 
The unperturbed orbits have different characteristic frequencies in different 
directions, but this would suggest that the resonant coupling could well be
stronger (or weaker) in one direction than in another.

\section{Modeling Luminosity Profiles in Real Gal-axies}
\subsection{Basic strategy}
The objective here is to show that the physical effects discussed
above, seemingly the inevitable consequence of a
supermassive black hole binary in the center of a galaxy, could provide
a natural explanation of the fact that, in a number of galaxies that have
been observed using {\em WFPC 2} ({\em e.g.,} Lauer et al 2002), the projected
surface brightness distribution in a given direction is not a monotonically
decreasing function of distance from the center of the galaxy.

The computations described here are not completely realistic. As in Section 2,
they assume black holes of exactly equal mass
executing exactly circular orbits, and the computed
orbits of test stars are not fully self-consistent since one is neglecting
both changes in the form of the bulk potential as stars are
displaced from their original trajectories and the slow decay of the 
binary orbit. They {\em do,} however,
demonstrate that allowing for a binary of relatively small size, ${\sim}{\;}
10$ pc, comprised of black holes with mass ${\la}{\;}1\%$ the mass of the
galaxy, leads generically to luminosity dips of the form that have been
observed. Moreover, fine-tuning parameters within a reasonable
range of values allows for the possibility of a comparatively detailed
(albeit in general nonunique) fit to observations of specific galaxies.
\par\noindent
The basic programme is as follows:
\par\noindent
${\bullet}$ Generate $N$-body realisations of a spherical galaxy characterised
by an isotropic distribution of velocities and a Nuker (Lauer {\em et al}
1995) density profile ${\rho}(r)$ with specified parameter values.
\par\noindent
${\bullet}$ Insert a black hole binary with specified masses
$M_{1}=M_{2}=M$ and radius $r_{h}$. For `realistic' values of $M$ and $r_{h}$,
$M(r_{h})$ is typically small compared with the black hole mass, so that one
can assume, at least approximately, that the black holes are executing a
Keplerian orbit with frequency ${\omega}=\sqrt{M/4r_{h}^{3}}$.
\par\noindent
${\bullet}$ Next evolve the initial conditions in the {\em fixed}
time-dependent potential comprised of the Nuker potential plus the potential
of the orbiting black hole binary, and track the radial density distribution
${\rho}(r,t)$. 
\par\noindent
${\bullet}$ Finally, assuming that mass traces light, compute line-of-sight
integrals along the density distribution to obtain integrated surface
densities and, hence, surface brightness distributions as functions of time.

Although this approach does not pretend to be completely realistic, it would
not seem totally unreasonable to insert the binary
`by hand' without allowing for the dynamics whereby it has evolved into a
tightly bound orbit near the galactic center. When the binary
orbit is very large, it will have a comparatively minimal effect. Energy and
mass transport only becomes important at comparatively small radii, where
$M{\;}{\ge}{\;}M(r_{h})$, and again become unimportant when the radius becomes
too small. Most of the action happens for a relativley limited range of radii.

Note, moreover, that the assumption $M{\;}{\ga}{\;}M(r_{h})$ tends to mitigate
the fact that the computations are not fully self-consistent: Although the
bulk forces associated with the galaxy cannot be neglected at all radii where
the binary has an appreciable effect, they {\em can} presumably be neglected,
at least approximately, at the comparatively small radii near the binary where
the effect of the black holes is strongest.

\subsection{The initial form of the density and potential}
Initial attempts at modeling using a spherical Dehnen potential
yielded results 
in qualitative agreement with observations.
However, comparatively large systematic deviations {\em were} observed, which 
appeared to
reflect the fact that the transition between the inner and outer power-law
profiles predicted by a Dehnen potential is too gradual to represent real
galaxies. For this reason, models were constructed instead using an initial
density distribution satisfying the more general Nuker Law (Lauer et al 1995)
\begin{equation}
{\rho}_{0}(r)={\rho}_{c}r^{-{\gamma}}
(1+r^{\alpha})^{({\gamma}-{\beta})\over {\alpha}}.
\end{equation}
Dehnen models are recovered for ${\alpha}=1$ and ${\beta}=4$. The central
density ${\rho}_{c}$ was chosen so that the total galactic mass $M_{g}=1.0$.
The associated potential $V(r)$ satisfies (in units with $G=1$)
\begin{equation}
V(r)=-4{\pi} \left[
{1\over r} \int_{0}^{r} {\rho}({\tilde r})\, {\tilde r}^{2}\,d{\tilde r}
+ \int_{r}^{\infty} {\rho}({\tilde r})\,{\tilde r}\,d{\tilde r} \right].
\end{equation}

Unfortunately, this potential can be expressed in terms of elementary
functions only for certain choices of ${\alpha}$ and ${\beta}$, which means,
generically, that orbits must be computed using an expensive interpolation
scheme. This motivated an effort to seek fits assuming values of ${\alpha}$
and ${\beta}$ for which $V$ {\em can} be expressed analytically. For the small
number of profiles considered hitherto, reasonable fits were achieved for
${\alpha}=2$ and ${\beta}=4$, which, for ${\gamma}=0$, 
yields a potential
\begin{equation}
V(r)=-{2\over {\pi}} {\tan^{-1} r\over r}
\end{equation}
and
\begin{equation}
M(r)={2\over {\pi}} \left(
\tan^{-1} r - {r\over 1+r^{2}} \right).
\end{equation}

Most models were constructed assuming $M(r_{h}){\;}{\ll}{\;}$ $M_{1}+M_{2}$,
so that the approximation of a Keplerian frequency is typically very good.
However, in an effort to allow for the influence of the galactic potential,
the models allowed for a slightly modified frequency 
${\omega}=\sqrt{{\cal M}/(2r_{h})^{3}}$,
where ${\cal M}=M_{1}+M_{2}+4M(r_{h})$. 

\subsection{Generating a surface brightness distribution}
Configuration space was divided into $N=100$ equally spaced concentric shells
$i$. Each shell corresponded to a range of energies, $E_{i-1}<E<E_{i}$,
$i=1,...,N$, sampled along the principal axes in the plane of the binary, but
perpendicular to the line connecting them. This was done to ensure that energy
was a monotonic function of radius, so that shuffling of energies could be
related directly to a redistribution of orbits in configuration space.
Each shell was sampled to select $M=300$ initial conditions, which were 
integrated for a time $t=512$. Orbital data were recorded
periodically and the new energies used to
reassign orbits to (in general) new shells. If $M_{i}(t)$ denotes the
number of orbits in shell $i$ at time $t$, then
\begin{equation}
{\Delta}_{i}(t)={M_{i}(t)\over M},
\end{equation}
the relative fluctuation in number, can be interpreted as a discretised
version of a radial density fluctuation ${\delta}(t)$ satisfying
\begin{equation}
{\rho}(r,t)=\left[ 1 + {\delta}(r,t)\right] {\rho}_{0}(r),
\end{equation}
with ${\rho}_{0}$ the initial density. ${\delta}(t)$
was interpolated from ${\Delta}_{i}(t)$ using a smooth-curve fitting routine.

The resulting density ${\rho}(r,t)$ was then integrated along the line of
sight to generate a surface brightness
\begin{equation}
{\mu}(r,t)={2\over \Upsilon} \int_{r}^{\infty}
{{\rho}({\tilde r},t){\tilde r}\over
\sqrt{{\tilde r}^{2}-r^{2}}}\, d{\tilde r}.
\end{equation}
Here ${\Upsilon}$ denotes the mass-to-light ratio, which was assumed constant
for the modeling described here.

\subsection{Results}
Figure 13 exhibits data for a typical model, corresponding to a Nuker Law
with ${\alpha}=2$, ${\beta}=4$, and ${\gamma}=0$. The binary parameters are
$M=0.005$, $r_{h}=0.15$, and ${\omega}=0.6086$.
The half-mass radius is $r=2.264$; $75\%$ of the mass is contained with $r=5$.

It is obvious that 
the binary induces a distinctive signature, characterised by an inversion
in both the mass density and the surface brightness profile. In some cases,
especially when ${\gamma}\;{\ne}\;0$, the contents of the innermost shells
can remain essentially intact. Aside, however, from those innermost shells,
one can identify a well-defined sphere of influence where the binary has
observable effects. For $r<r_{1}$, there is a systematic underpopulation
of stars and, hence, a dip in luminosity; for $r_{1}<r<r_{2}$, there is 
a systematic overpopulation or bulge resulting from stars transported outwards 
from radii $r<r_{2}$. For $r>r_{2}$ the density and surface brightness
distribution remain essentially unchanged. Significantly, the signature, 
once established, remains largely unchanged in bulk properties. In particular, 
the dip is comparably prominent visually at $t=128$ and $t=512$.

Figure 14 exhibits a more systematic attempt to model the
surface brightness of {\em NGC} 3706, again starting from a Nuker Law with
${\alpha}=2$, ${\beta}=4$, and ${\gamma}=0$. Physical distance was translated
into angular separation assuming a scaling such that
$r=1$ corresponds to 0.15 arcsec or, given the distance estimate given by
Lauer {\em et al}, $r{\;}{\approx}{\;}24$ pc.
Once again $M=0.005$, but now $r_{h}=0.085$, which
corresponds to a physical $r_{h}{\;}{\sim}{\;}2.0$ pc and an angular 
separation ${\sim}{\;}0.014$ arcsec. The orbital frequency ${\omega}=1.567$,
which implies an orbital period ${\tau}{\;}{\approx}{\;}4.00$.

In this case, the inner dip extends out to ${\sim}{\;}0.10$ arcsec; the bulge
extends to  ${\sim}{\;}0.30$ arcsec. On scales ${\ga}{\;}0.30$ arcsec, 
{\em i.e.,} $r{\;}{\ga}{\;}75$ pc, the binary has only a comparatively minimal 
effect, so that the surface brightness remains essentially unchanged. 

The perturbed Nuker
model is quite successful in modeling the dip and the outer region, where 
errors in surface brightness correspond typically to 
${\delta}{\mu}{\;}{\sim}{\;}0.005$ magnitudes or less. In particular, 
the dip is much better fit by the perturbed model than by 
an unperturbed Nuker
model. Both qualitatively -- in that an unperturbed Nuker model
requires a monotonically decreasing surface brightness -- and quantitatively
-- in terms of the actual error ${\delta}{\mu}$ --, the perturbed model does
a much better job. However, both the perturbed and unperturbed models are
somewhat less successful in accounting for the detailed shape of the bulge
(although the model with a binary does a bit better).

There are at least two possible explanations for this lack of success. Most
obvious is the fact that, demanding ${\alpha}=2$ and ${\beta}=4$, so that
the potential could be written in terms of elementary functions, limits
one's flexibility in modeling the transition region between the inner
and outer (unperturbed) power law profiles. Allowing for fractional parameter
values (which requires that the potential be computed
numerically) will likely yield better fits. However, it is also possible that
this lack of success reflects in part the oversimplistic character of this
kinematic model.
In a real galaxy, the binary orbit decays as the binary tranfers energy to
the stars; and the fact that $r_{h}$ is not really constant might
be expected to have some observable effects. Attempts to remedy these
deficiencies of the model are currently underway.

It should, however, be stressed that the general effects of the 
binary are relatively insensitive to $r_{h}$, provided only that $M>M(r_{h})$.
This is, {\em e.g.,} evident from Fig.~15, which exhibits surface brightness 
distributions at $t=256$ for both the model considered in Fig.~14 and another 
model identical except that $r_{h}=0.025$, a radius only $0.29$ times as 
large. 
There are some differences in detail, but neither fit is clearly
superior visually. It is also evident from Fig.~14 that the basic observable 
structure develops very
quickly. Although the details of the surface brightness profile can vary
considerably for times as long as $t{\;}{\sim}{\;}128$ or more, the existence 
of the dip region is obvious already by $t{\;}{\sim}{\;}32$, about $8$ binary 
orbital periods for the $r_{h}=0.085$ model.

Although the model described here is kinematic, one can try to describe how it 
might be manifested in a self-consistent description: When the binary orbit
is too large, its decay will be dominated by more conventional processes,
discussed, {\em e.g.,} in Tremaine \& Weinberg (1984) and Nelson \& Tremaine 
(1999). However, once the radius is sufficiently small that 
$M(r_{h}){\;}{\sim}{\;}2M$, the resonant phase mixing described here -- which
can be viewed as a variant of Tremaine's resonant relaxation -- will be 
triggered. As additional energy is transferred to the stars, the binary will 
continue to decay and, when the size of the orbit becomes too small, the 
effect will again `turn off.'


For the models in Fig.~14 and 15, 
the process should `turn on' at $r_{h,1}{\;}{\approx}{\;}0.2$ and
`turn off' at $r_{h,2}{\;}{\approx}$
${\;}0.005$. However, during this interval, the
binary will have lost an energy ${\sim}{\;}M^{2}/r_{h,2}{\;}{\sim}{\;}0.02$,
several percent of 
the energy of the galaxy at the time that the 
process begins. The obvious questions, therefore, are: How long does
it take for a binary with radii $r_{h}$ satisfying 
$r_{h,2}{\;}{\la}{\;}r_{h}{\;}{\la}r_{h,1}$ to pump this much energy into 
the stars? And is this long enough to establish the signature observed in
Figs.~14 and 15? The time required depends to
a certain extent on the precise value of $r_{h}$. However, an analysis of
the models in Figs.~14 and 15, as well as models with somewhat larger and
smaller values of $r_{h}$ indicates that the total energy required to 
establish the observed signature is relatively small. For example, the model
with $r_{h}=0.085$ entailed an increase in galactic energy of order $1\%$ 
at $t=32$ and $3\%$ at $t=512$. The model with $r_{h}=0.025$ yielded $1.5\%$
at $t=32$ and $6\%$ at $t=512$.

Alternatively, a binary decay rate can be estimated as follows: Given that
the pumping of energies into the stars is diffusive, the decay time ${\tau}$
should satisfy
${\tau}/T{\;}{\sim}{\;}(M^{2}/r_{h}{\langle}{\delta}E{\rangle})^{2},$
where $T$ is the time over which ${\langle}{\delta}E{\rangle}$ is computed.
Supposing, however, that $r_{h}{\;}{\sim}{\;}0.01$, $M(r_{h}){\;}{\sim}{\;}
0.01$, ${\langle}{\delta}E{\rangle}{\;}{\sim}{\;}0.01$, and 
$T{\;}{\sim}{\;}10$, one infers that ${\tau}{\;}{\sim}{\;}T{\;}{\sim}{\;}10$.
For the orbit to shrink from $r_{h}=0.2$ to $r_{h}=0.005$, a factor of $40$,
would require a few ${\tau}$, say $t{\;}{\sim}{\;}50$, an interval long enough
to establish a distinctive luminosity dip.

One final point should be noted. 
Attributing such a luminosity dip to a supermassive binary does not
necessarily imply that the binary should still be present. 
If, neglecting the binary, the galaxy can be idealised as a collisionless
equilibrium, one might expect that a dip in surface brightness, once
generated, could persist even after the binary
has coalesced, at least for times short compared with the time scale on
which stars at larger radii could be scattered inwards via collisional
relaxation. To the extent that the bulk potential is time-independent, 
in the absence of `collisions' energy is conserved, so that an 
underpopulated region in energy space cannot be repopulated.

\section{Discussion}
The computations described here yield several significant conclusions about 
phase mixing in a time-dependent
potential. Most obvious is the fact that a supermassive black hole
binary can serve as an important source of transient chaos which facilitates
efficient resonant phase mixing, shuffling the energies of stars (or any
other objects) as well as phase space coordinates on a constant energy
hypersurface. In particular, the effects observed here from a comparatively
`small scale' perturbation are quite similar to the effects
observed when galaxies are subjected to larger scale systematic oscillations
(Kandrup, Vass \& Sideris 2003). It is especially striking that,
even though the perturbation is relatively low amplitude and concentrated
very near the center of the galaxy, it can have significant effects at
comparatively large radii. All this reinforces the expectation
that resonant phase mixing could be a generic physical effect in
galaxies subjected to an oscillatory time dependence.

Contrary, perhaps, to naive expectation, it appears that the shuffling of
energies is diffusive, rather than exponential, so that energy phase
mixing is less dramatic than phase mixing of coordinates and velocities.
Even though the time-dependent perturbation can increase both the relative
abundance of chaotic orbits and the degree of exponential sensitivity exhibited
by chaotic orbits, it need not make orbits
exponentially unstable in the phase space direction orthogonal to the constant
energy hypersurfaces.

However, such energy shuffling could still play
an important role in violent relaxation. Indeed,
the fact that this energy shuffling is not exponential is consistent
with self-consistent simulations of violent relaxation ({\em e.g.,}
Quinn \& Zurek 1988) which indicate that,
even though `particles' 
are almost completely
`randomised' in terms of most phase space coordinates, they exhibit
a partial remembrance of initial conditions. In particular,
`particles' that start with low (high) binding energies tend systematically
to end with low (high) binding energies. If, {\em e.g.,} stars in simulations
involving hard, head-on collisions of galaxies are ordered in terms of their
initial and final binding energies, the rank correlation ${\cal R}$ between
the initial and final ordered lists typically satisfies
(Kandrup, Mahon \& Smith 1993) ${\cal R}{\;}{\ga}{\;}0.6$.

That a supermassive binary will cause a systematic readjustment
in the density distribution of the host galaxy seems largely independent of
the form of the galactic potential or the orbital parameters of the binary,
although the precise form of the readjustment does depend on these details.
In particular, one sees qualitatively similar effects for Dehnen potentials
with different cusp indices ${\gamma}$ and for Nuker Laws with different
transitional radii properties. Similarly, the eccentricity and the orientation
of the supermassive binary are not crucial, and allowing for unequal, but still
comparable, masses does not result in qualitative changes. Irrespective of
all these details, when the total binary mass
$M_{1}+M_{2}{\;}{\ll}{\;}M(r_{h})$, with $r_{h}$ the `size' of the binary
orbit, stars cannot resonate with the binary and comparatively little mass
transport occurs. However, when $M_{1}+M_{2}{\;}{\sim}{\;}M(r_{h})$, one
starts seeing substantial effects which can extend to radii
${\gg}{\;}r_{h}$.

One might therefore expect that, when its orbit is large, the
binary will have only a minimal effect on the bulk properties of the galaxy;
but that when, as a result of dynamical friction ({\em e.g.,} Merritt 2001),
the orbit has decayed to a sufficiently small size, it will begin to have an
appreciable -- and observable -- effect.
\vskip .2in
HEK acknowledges useful discussions with Christos Siopis, who tried to
convince him of the importance of explaining luminosity `dips' months before
he was ready to listen.
HEK, IVS, and BT were supported in part by NSF AST-0070809. IVS and CLB were
supported in part by Department of Education grant G1A62056. We would like to 
thank the Florida State University School of Computational Science and 
Information Technology for granting access to their supercomputer facilities.
\vskip 0.5in

\centerline{REFERENCES}
\vskip .15in
\par\noindent
Bertin, G. 2000, Dynamics of Galaxies, Cambridge University Press,
Cambridge
\par\noindent
Binney, J. 1978, Comments Astrophys., 8, 27
\par\noindent
Dehnen, W. 1993, MNRAS, 265, 250
\par\noindent
Kandrup, H. E., 1998, MNRAS, 301, 960
\par\noindent
Kandrup, H. E., 2003, in: Springer Lecture Notes in Physics,
in press (astro-ph/0212031)
\par\noindent
Kandrup, H. E., Mahon, M. E. 1994, Phys. Rev. E 49, 3735
\par\noindent
Kandrup, H. E., Mahon, M. E., Smith, H. 1993, A\&A, 271, 440
\par\noindent
Kandrup, H. E., Siopis, C. 2003, MNRAS, submitted (astro-ph/0305198)
\par\noindent
Kandrup, H. E., Terzi{\'c}, B. 2003, MNRAS, submitted
\par\noindent
Kandrup, H. E., Vass, I. M., Sideris, I. V. 2003, MNRAS, 341, 927
\par\noindent
Kormendy, J., Bender, R. 1996, ApJ Lett., 464, 119
\par\noindent
Lauer, T. et al, 1995, AJ, 110, 2622
\par\noindent
Lauer, T. et al, 2002, AJ, 124, 1975
\par\noindent
Lichtenberg, A. J., Lieberman, M. A. 1992, Regular and Chaotic
Dynamics, Springer, New York.
\par\noindent
Lynden-Bell, D. 1967, MNRAS, 136, 101
\par\noindent
Merritt, D. 2001, ApJ, 556, 445
\par\noindent
Merritt, D., Fridman, T. 1996, ApJ, 460, 136
\par\noindent
Merritt, D., Valluri, M. 1996, ApJ, 471, 82
\par\noindent
Nelson, R. W., Tremaine, S. 1999, MNRAS, 306, 1
\par\noindent
Quinn, P. J., Zurek, W. H. 1988, ApJ, 331, 1
\par\noindent
Tremaine, S., Weinberg, M. D. 1984, MNRAS, 209, 729
\vfill\eject
\clearpage
\begin{figure}
\plotone{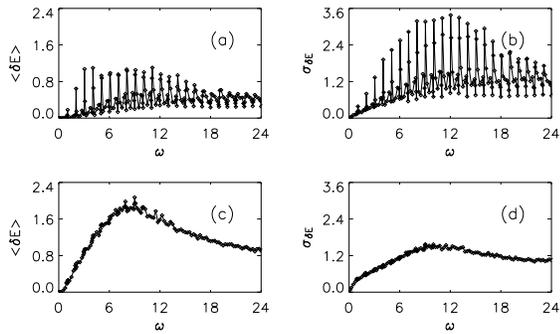}
\caption{(a) The mean shift in energy ${\langle}{\delta}E{\rangle}$ for all 
the orbits in a 1600 orbit ensemble with $E=0.87$ and 
${\langle}r_{in}{\rangle}{\;}{\approx}{\;}0.86$, evolved in a spherical 
oscillator potential with $M=0.05$, $r_h=0.3$, and $a^{2}=b^{2}=c^{2}=1.0$, 
plotted as a function of frequency ${\omega}$.
(b). The dispersion ${\sigma}_{{\delta}E}$ for all the orbits.
(c) - (d) The same as the preceding for orbits integrated in a
potential with $a^{2}=1.33$, $b^{2}=1.0$, and $c^{2}=0.80$ and an ensemble
with $E=0.87$ and ${\langle}r_{in}{\rangle}{\;}{\approx}{\;}0.89$.}
\end{figure}

\begin{figure}
\plotone{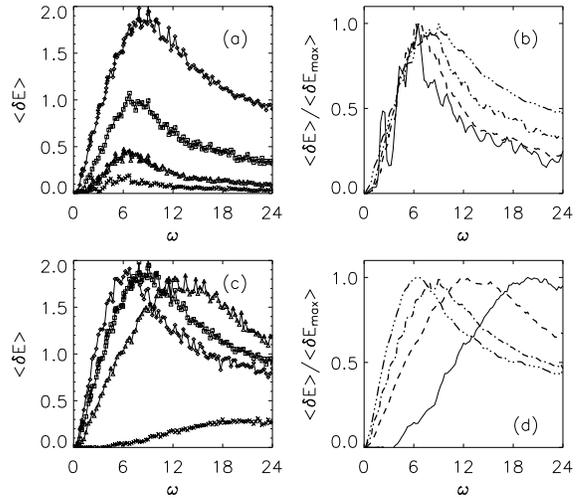}
\caption{(a) The mean energy shift ${\langle}{\delta}E{\rangle}$ for the 
same ensemble used to generate FIG.~1 (c) and (d), integrated with  $r_h=0.3$,
$a^{2}=1.33$, $b^{2}=1.0$, $c^{2}=0.80$, and (from top to bottom) 
$M=0.05$, $M=0.0281$, $M=0.0158$, and $M=0.005$. (b) 
${\langle}{\delta}E{\rangle}$ for the same ensembles -- solid line for 
$M=0.005$, dashed line for $M=0.0158$, dot-dashed line for $M=0.0281$, and
triple-dot-dashed for $M=0.05$ -- now expressed in units of the maximum shift
${\langle}{\delta}E_{max}{\rangle}$.
(c)  The mean energy shift ${\langle}{\delta}E{\rangle}$ for integrations with
$a^{2}=1.33$, $b^{2}=1.0$, $c^{2}=0.80$, $M=0.05$, and (curves peaking from 
left to right) $r_h=0.4$. $r_h=0.3$, $r_h=0.2$, and $r_h=0.1$.
(d) ${\langle}{\delta}E{\rangle}$ expressed in units of the maximum shift
${\langle}{\delta}E_{max}{\rangle}$ -- solid line for $r_h=0.1$, dashed for
$r_h=0.2$, dot-dashed for $r_h=0.3$, and triple-dot-dashed for $r_{h}=0.4$.}
\end{figure}

\begin{figure}
\plotone{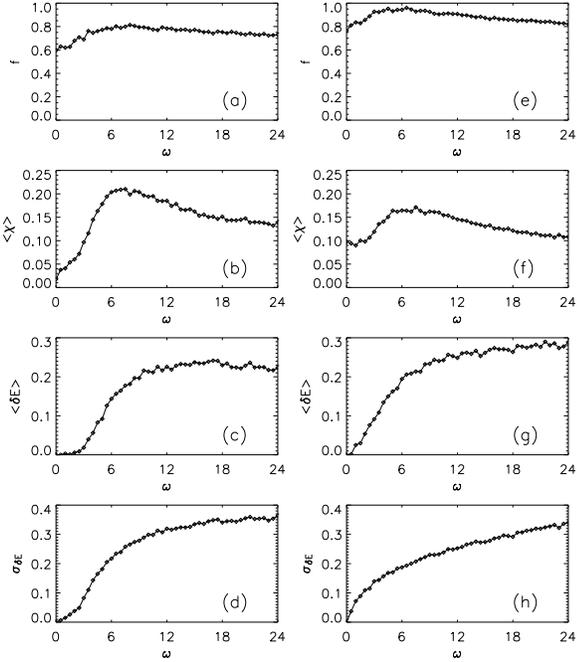}
\caption{(a) The fraction $f$ of strongly chaotic orbits in a
1600 orbit ensemble with initial energy $E=-0.70$ and mean initial radius 
${\langle}r_{in}{\rangle}{\;}{\approx}{\;}0.33$,
evolved in a spherically symmetric Dehnen potential with ${\gamma}=1.0$,
$M=0.01$, $r=0.05$, and
$a^{2}=b^{2}=c^{2}=1.00$. 
(b) The mean value ${\langle}{\chi}{\rangle}$ of the largest finite time 
Lyapunov exponent for the strongly chaotic orbits.
(c) The mean shift in energy ${\langle}{\delta}E{\rangle}$ for all the orbits.
(d) The dispersion ${\sigma}_{{\delta}E}{\rangle}$ for all the orbits.
(e) - (h) The same as the preceding for orbits integrated in a
potential with $a^{2}=1.25$, $b^{2}=1.00$, and $c^{2}=0.75$, again for an 
ensemble with $E=-0.70$ and ${\langle}r_{in}{\rangle}{\;}{\approx}{\;}0.33$.}
\end{figure}

\begin{figure}
\plotone{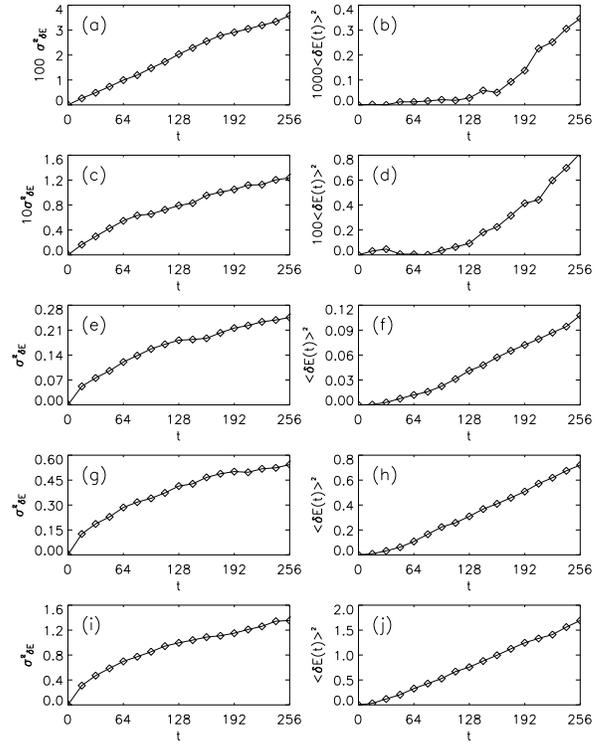}
\caption{(a) ${\sigma}^{2}_{{\delta}E}$, where ${\sigma}_{{\delta}E}(t)$ is
the time-dependent spread in energy shifts associated with an ensemble of 
orbits evolved in an oscillator potential 
with $M=0.05$, $r_{h}=0.3$, $a^{2}=1.33$, $b^{2}=1.0$, $c^{2}=0.80$ and
${\omega}=0.5$. (b) ${\langle}{\delta}E(t){\rangle}^{2}$ for the same ensemble.
(c) and (d) The same for ${\omega}=1.0$. 
(e) and (f) The same for ${\omega}=2.0$. 
(g) and (h) The same for ${\omega}=4.0$. 
(i) and (j) The same for ${\omega}=8.0$.}
\end{figure}

\begin{figure}
\plotone{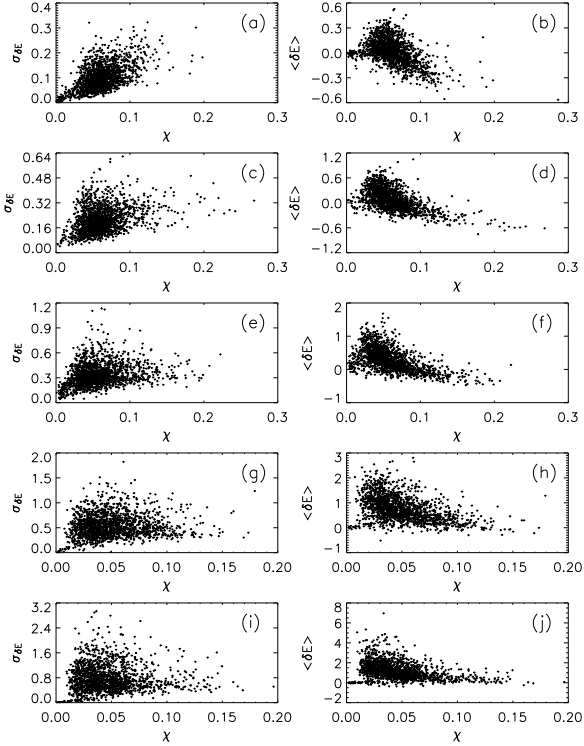}
\caption{(a) Scatter plots relating ${\sigma}_{{\delta}E}$ and ${\chi}$,
where ${\sigma}_{{\delta}E}$ represents the dispersion associated with the
time-dependent ${\delta}E(t)$ {\em for an individual orbit} over the interval
$0<t<512$. The orbits are the same that were used to generate FIG.~3, 
integrated with 
${\omega}=0.5$. (b) Scatter plots relating ${\langle}{\delta}E{\rangle}$
and ${\chi}$, where  ${\langle}{\delta}E{\rangle}$ represents the mean
value of  ${\delta}E(t)$, computed for the same orbits as in (a).
(c) and (d) The same for ${\omega}=1.0$. 
(e) and (f) The same for ${\omega}=2.0$. 
(g) and (h) The same for ${\omega}=4.0$. 
(i) and (j) The same for ${\omega}=8.0$. }
\end{figure}

\begin{figure}
\plotone{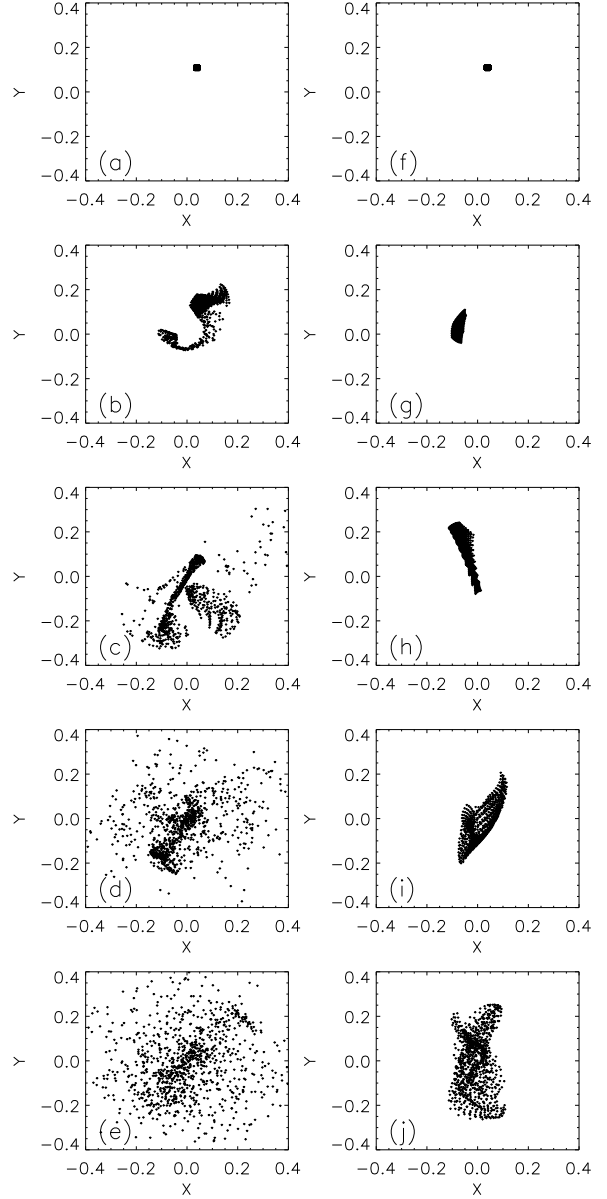}
\caption{(a) The $x$ and $y$ coordinates at $t=0$ for an initially localised 
ensemble of orbits with $E=-0.70$ and 
${\langle}r_{in}{\rangle}{\;}{\approx}{\;}0.33$,
evolved in a spherical Dehnen potential with 
${\gamma}=1.0$, $r_{h}=0.05$, $M=0.005$, and ${\omega}=\sqrt{10}$.
(b) $t=8$. (c) $t=16$. (d) $t=32$. (e) $t=64$.
(f) - (j). The same for stationary black holes, {\em i.e.,} ${\omega}=0.0$. }
\end{figure}

\begin{figure}
\plotone{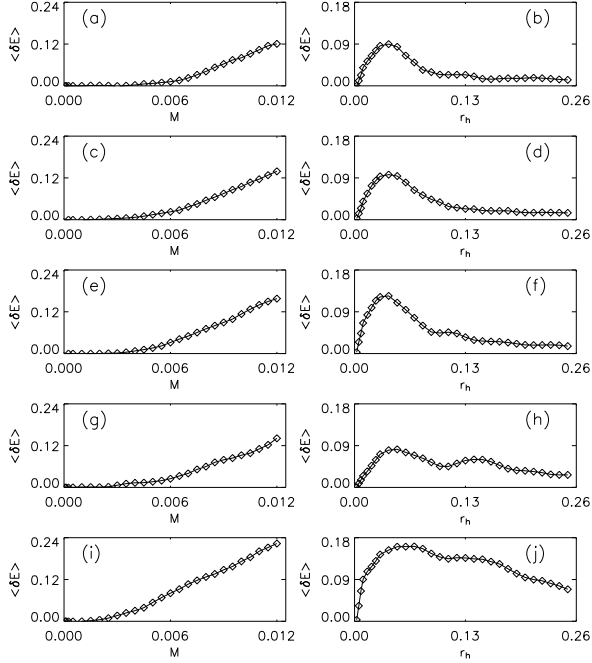}
\caption{(a) The mean shift in energy, ${\langle}{\delta}E{\rangle}$, computed
for an ensemble of orbits with $E=-0.70$ and
${\langle}r{\rangle}{\;}{\approx}{\;}0.33$, evolved in a ${\gamma}=1.0$
Dehnen model with $a^2=b^2=c^2=1$ in the presence of a supermassive binary
comprised of two black holes executing
strictly circular orbits with $r_{h}=0.05$ and different values of
$M_{1}=M_{2}{\;}{\equiv}{\;}M$. (b)  ${\langle}{\delta}E{\rangle}$ for the
same ensemble
evolved in the same Dehnen model, again allowing for a binary executing
circular orbits, but now with $M_{1}=M_{2}=0.01$ and variable $r_{h}$.
(c) and (d) The same for a model with $a^2=b^2=0.90$ and $c^{2}=1.21$.
(e) and (f) The same for a model with $a^2=b^2=1.21$ and $c^{2}=0.64$
(g) and (h) The same for a model with $a^2=1.10$, $b^2=1.0$ and $c^{2}=0.90$.
(i) and (j) The same for a model with $a^2=1.25$, $b^2=1.0$ and $c^{2}=0.75$.
In each case, the frequency ${\omega}=\sqrt{(M_{1}+M_{2})/a^{3}}$, with
$A$ the semi-major axis.
}
\end{figure}

\begin{figure}
\plotone{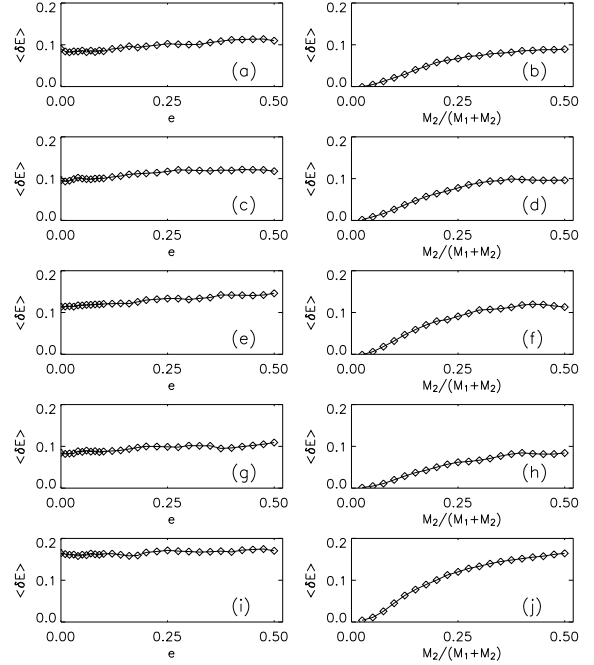}
\caption{(a) The mean shift in energy, ${\langle}{\delta}E{\rangle}$, computed
for an ensemble of orbits with $E=-0.70$ and
${\langle}r{\rangle}{\;}{\approx}{\;}0.33$, evolved in a ${\gamma}=1.0$
Dehnen model with $a^2=b^2=c^2=1$ in the presence of a supermassive binary
comprised of two black holes with mass $M_{1}=M_{2}=0.01$ executing orbits with
semi-major axis $A=0.10$ and variable eccentricity $e$.
(b)  ${\langle}{\delta}E{\rangle}$ for the same ensemble evolved in the same 
Dehnen model, again assuming $M_{1}+M_{2}=0.02$ and $a=0.10$, but now allowing 
for different ratios $M_{2}/(M_{1}+M_{2})$.
(c) and (d) The same for a model with $a^2=b^2=0.90$ and $c^{2}=1.21$.
(e) and (f) The same for a model with $a^2=b^2=1.21$ and $c^{2}=0.64$
(g) and (h) The same for a model with $a^2=1.10$, $b^2=1.0$ and $c^{2}=0.90$.
(i) and (j) The same for a model with $a^2=1.25$, $b^2=1.0$ and $c^{2}=0.75$.
}
\end{figure}

\begin{figure}
\plotone{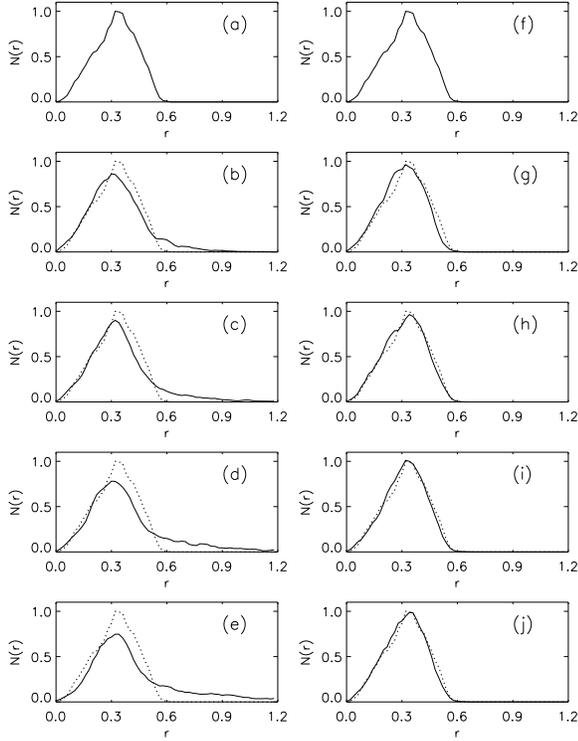}
\caption{(a) The initial angle-averaged radial density distribution 
associated with a 4800 orbit sampling of the $E=-0.70$ constant energy
hypersurface, subsequently integrated in a pseudo-Dehnen potential with 
${\gamma}=1.0$, 
$M=0.01$, $r_{h}=0.05$, $a^{2}=1.25$, $b^{2}=1.00$, $c^{2}=0.75$ and ${\omega}=
\sqrt{20}$.
(b) The density at $t=16$. (The dotted line reproduces the initial 
distribution.) (c) $t=32$. (d) $t=64$. (e) $t=128$.
(f) - (j) The same for stationary black holes, {\em i.e.,} ${\omega}=0.0$.}
\end{figure}

\begin{figure}
\plotone{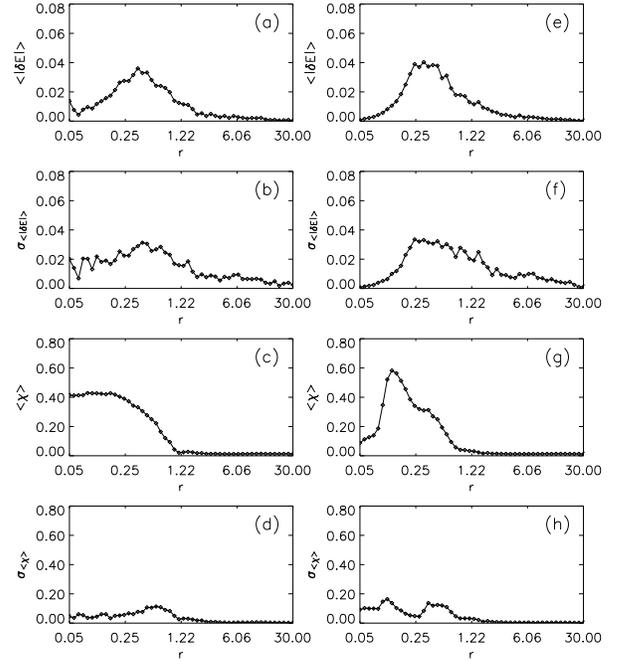}
\caption{(a) Mean energy shift ${\langle}{\delta}E{\rangle}$, computed for
ensembles with different radii, for orbits in a spherical Dehnen model with
${\gamma}=0.0$ and $a=b=c=1.0$ and black hole parameters $M=0.005$, 
$r_{h}=0.25$,
and ${\omega}=0.2828$. Note that the radius $r$ is plotted on a logarithmic
scale. (b) The dispersion ${\sigma}_{{\delta}E}$ for the
same ensembles. (c) The mean value ${\langle}{\chi}{\rangle}$ for each 
ensemble. (d) The dispersion ${\sigma}_{\chi}$. (e) - (h) The same for a
model with ${\gamma}=1.0$.}
\end{figure}

\begin{figure}
\plotone{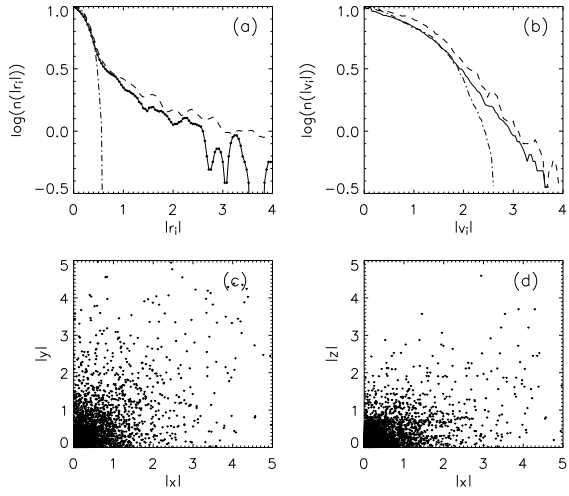}
\caption{(a) The direction-dependent spatial distributions $n(|x|)$ (solid
curve) and $n(|z|)$ (dashes) at $t=512$ generated for a $4800$ orbit sampling
of the $E=-0.70$ hypersurface with ${\gamma}=1.0$, $M=0.01$, $r_{h}=0.005$, 
$a=b=c=1$, and ${\omega}=\sqrt{20}$,
along with the distribution $n(|x|)$ (dot-dashed) at time $t=0$.
(b) The corresponding direction-dependent velocity distributions.
(c) $x$ and $y$ coordinates for the ensemble at $t=512$. 
(d) $x$ and $z$ coordinates for the ensemble at $t=512$. }
\end{figure}

\begin{figure}
\plotone{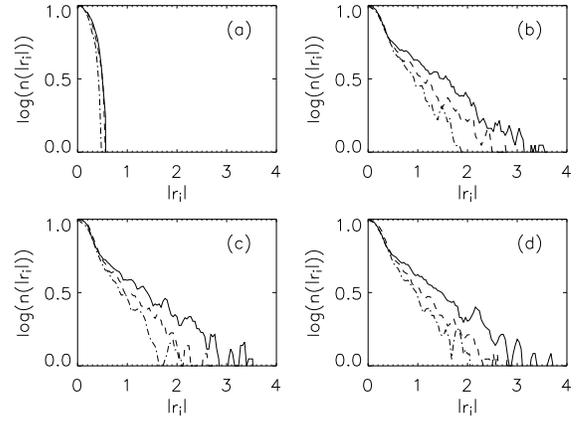}
\caption{Direction-dependent spatial distributions $n(|x|)$ (solid
curve), $n(|y|)$ (dashes), and $n(|z|)$ (dot-dashes) generated for a $4800$ 
orbit sampling of the $E=-0.62$ hypersurface with ${\gamma}=1.0$, $M=0.01$, 
$r_{h}=0.005$, 
$a^{2}=1.25$, $b^{2}=1.0$, $c^{2}=0.75$, and ${\omega}=\sqrt{20}$.
along with the distribution $n(|x|)$ (dot-dashed) at time $t=0$.
(a) The distributions at time $t=0$. (b) The distributions at $t=512$, allowing
for a binary orbiting in the $x-y$ plane. (c) The same for a binary in the
$y-z$ plane. (d) The $z-x$ plane.}
\end{figure}

\begin{figure}
\plotone{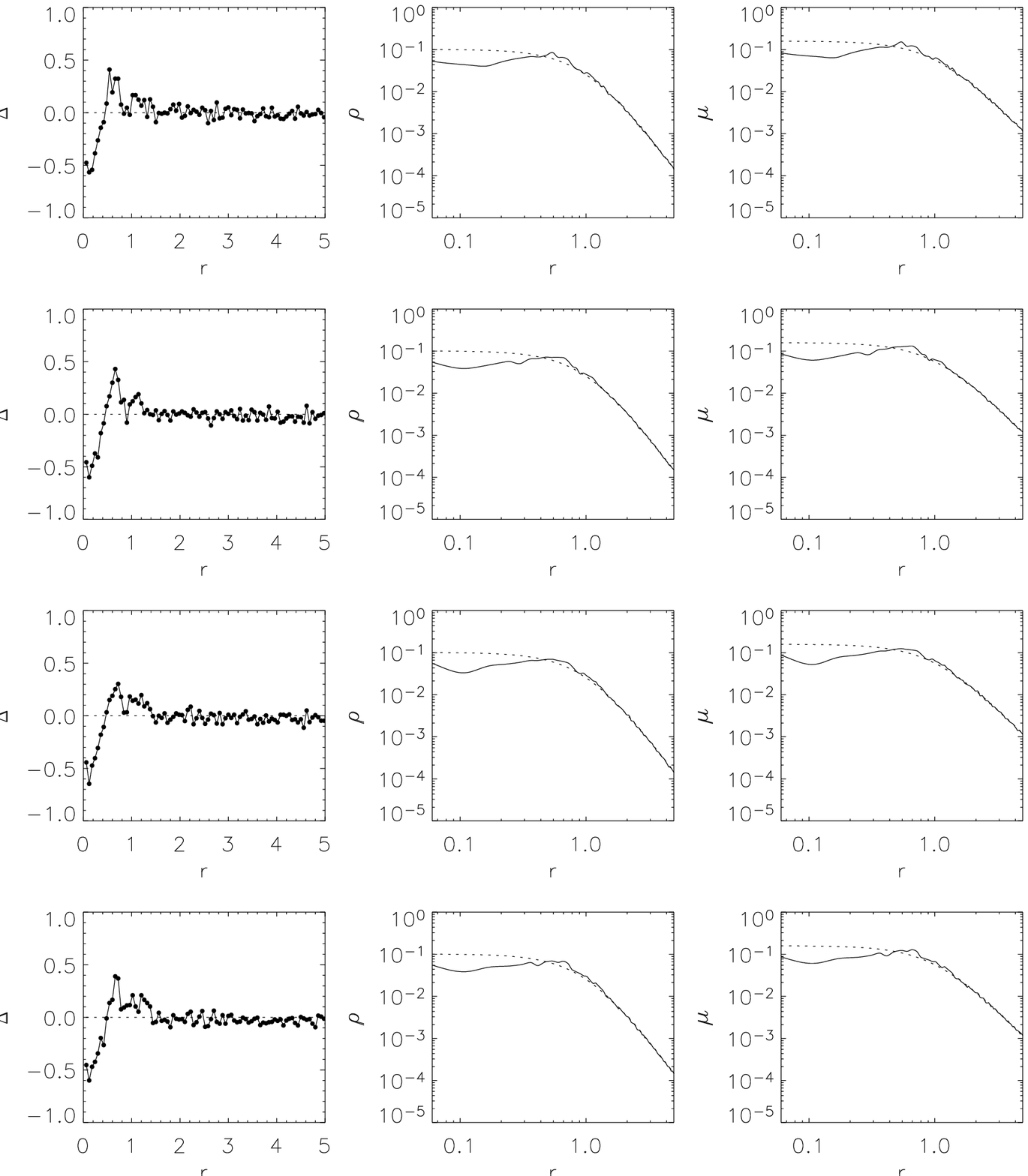}
\caption{Computed quantities for a Nuker model with ${\alpha}=2$,
${\beta}=4$, ${\gamma}=0$, $M_{BH}=0.005$, $r_{h}=0.15$, 
and ${\omega}=0.6086$. The first column exhibits ${\Delta}$, the relative 
fluctuation in density for different shells; the second exhibits the
interpolated smooth density ${\rho}$; the third exhibits the surface 
brightness, assuming that mass traces light. From top to bottom, rows 
represent integration times $t=128$, $256$, $384$, and $512$. In each case,
the dotted lines represent the original unperturbed values.}
\end{figure}

\begin{figure}
\plotone{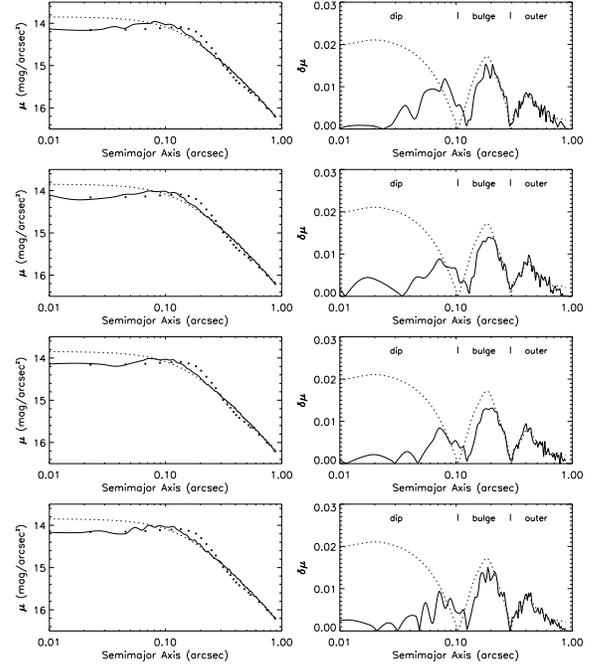}
\caption{Modeling {\em NGC} 3706 with a Nuker model with ${\alpha}=2$, 
${\beta}=4$, ${\gamma}=0$, $r_{h}=0.085$, and ${\omega}=1.567$. 
The left column exhibits the observed surface brightness profile (solid 
circles), the surface density predicted by an unperturbed Nuker Law (dotted
lines), and the time-dependent surface density generated by the binary
(solid lines) at times (from top to bottom) $t=32$, $t=64$, $t=128$, and 
$t=256$. The right column exhibits the relative error of the
the fit for a Nuker model without (dashes) and with the binary (solid lines).
}
\end{figure}

\begin{figure}
\plotone{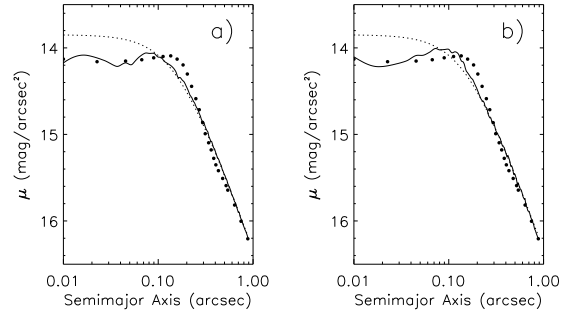}
\caption{(a) The best fit model with ${\alpha}=2$, 
${\beta}=4$, ${\gamma}=0$, $r_{h}=0.025$, and ${\omega}=8.968$ at time $t=256$.
(b) The same, except assuming $r_{h}=0.085$ and ${\omega}=1.567$.
}
\end{figure}
\end{document}